\documentclass[preprint2]{aastex}
\usepackage{rotating}
\usepackage{graphicx}
\usepackage{grffile}
\usepackage{threeparttable}

\begin{document}
\setcounter{secnumdepth}{2}
\title{The TRENDS High-Contrast Imaging Survey. VII. \\ Discovery of a Nearby Sirius-like White Dwarf System (HD~169889)}
\author{Justin R. Crepp\altaffilmark{1}, Erica J. Gonzales\altaffilmark{1,2,9}, Brendan P. Bowler\altaffilmark{3,7}, Farisa Morales\altaffilmark{4}, Jordan Stone\altaffilmark{5}, Eckhart Spalding\altaffilmark{5}, Amali Vaz\altaffilmark{5,6}, Philip Hinz\altaffilmark{5}, Steve Ertel\altaffilmark{5}, Andrew Howard\altaffilmark{7}, Howard Isaacson\altaffilmark{8}}
\email{jcrepp@nd.edu} 
\altaffiltext{1}{Department of Physics, University of Notre Dame, 225 Nieuwland Science Hall, Notre Dame, IN, 46556, USA}
\altaffiltext{2}{Department of Astronomy, University of California, Santa Cruz, 1156 High Street, Santa Cruz, CA 95064}
\altaffiltext{3}{Department of Astronomy, The University of Texas at Austin, TX, 78712, USA} 
\altaffiltext{4}{Jet Propulsion Laboratory, 4800 Oak Grove Dr, Pasadena, CA 91109, USA}
\altaffiltext{5}{Steward Observatory, Department of Astronomy, University of Arizona, 933 N. Cherry Ave, Tucson, AZ 85721, USA}
\altaffiltext{6}{Large Binocular Telescope, Mount Graham International Observatory, Safford, AZ 85546}
\altaffiltext{7}{Department of Astronomy, California Institute of Technology, 1200 E. California Blvd., Pasadena, CA 91125}
\altaffiltext{8}{Department of Astronomy, University of California, Berkeley, CA 94720, USA} 
\altaffiltext{9}{NSF Graduate Research Fellow}

\begin{abstract}  
Monitoring the long-term radial velocity (RV) and acceleration of nearby stars has proven an effective method for directly detecting binary and substellar companions. Some fraction of nearby RV trend systems are expected to be comprised of compact objects that likewise induce a systemic Doppler signal. In this paper, we report the discovery of a white dwarf (WD) companion found to orbit the nearby ($\pi = 28.297 \pm 0.066$ mas) G9 V star HD~169889. High-contrast imaging observations using NIRC2 at Keck and LMIRCam at the LBT uncover the ($\Delta H = 9.76 \pm 0.16$, $\Delta L' = 9.60 \pm 0.03$) companion at an angular separation of 0.8'' (28 au). Thirteen years of precise Doppler observations reveal a steep linear acceleration in RV time series and place a dynamical constraint on the companion mass of $M \geq 0.369 \pm 0.010 M_{\odot}$. This ``Sirius-like" system adds to the census of WD companions suspected to be missing in the solar neighborhood.  
\end{abstract}                                                                                                                                                                                                                                                                                                                                                                                                                                                                                                                                                                                                                                                                                                                                                                                                                                                                                                                                                                                                                                      
\keywords{keywords: techniques: radial velocities, high angular resolution; astrometry; stars: individual (HD~169889, HIP~90365, SAO~123479), white dwarfs}   

\section{INTRODUCTION}\label{sec:intro}
% What is completeness of WD population? How are they normally detected? Why are there missing WD's?
Many undiscovered white dwarf (WD) companions remain hidden in the solar neighborhood, their faint signals overwhelmed by the glare of light emitted from a now-brighter unevolved companion star at optical and infrared wavelengths. The occurrence of these ``Sirius-like" systems --- binaries with a WD orbiting a solar-type or earlier-type star --- represents only $\approx$8\% of known WDs in close proximity to the Sun; this rate falls precipitously at distances beyond $\approx$25 pc due to observational bias \citep{holberg_13}. WD companions are routinely missed from seeing-limited surveys that are relegated to studying large projected separations (Parsons et al. 2016).

% Although expected to be uncommon based on theoretical grounds (having to do with the dynamical evolution and frictional forces involved in post common-envelope evolution)  

% Why are directly imaged benchmark WDs useful?
Directly imaged WDs in Sirius-like systems are valuable benchmark objects that confer information regarding: empirical mass--radius relations; the luminosity function of WDs from precise parallax measurements of the host star; calibration of progenitor mass relations; atmospheric characterization and chemical constituents; and cooling timescale based on age estimates of the host star \citep{parsons_17,harris_06,kilic_06,matthews_14,holberg_16,bacchus_17,joyce_17}. 

% Are there other ways of detecting benchmark WDs?
\emph{Hubble Space Telescope} imaging follow-up measurements based on the presence of excess UV flux has been shown as a reliable method to identify Sirius-like WD companions \citep{parsons_16}. At near-infrared wavelengths, ground-based adaptive optics (AO) observations of Sun-like stars can also be employed. Given that a blind survey (that is uninformed by dynamical signposts or other means) to detect WD companions would be inefficient, long-term radial velocity (RV) monitoring of Sun-like stars can instead be used to reveal the presence of compact objects as their orbits evolve over timescales of years to decades \citep{crepp_14,rodigas_16}. Orbital monitoring through imaging and precise RV measurements further permits studies of the period and eccentricity distribution of WD companions as currently only a handful have complete orbits \citep{bond_17}. The orbital architecture of WD binaries informs studies that relate orbital evolution to the loss of companions (the rate of which is predicted to be significant at short orbital periods), circumstellar disks, mass transfer, and type Ia supernovae explosions \citep{parsons_16}. 

% What is this paper about? 
In this paper, we report the discovery of a Sirius-like WD companion orbiting the nearby ($d=35.34 \pm 0.08$ pc) Sun-like star HD~169889. The companion was originally mistaken as a brown dwarf candidate based on its brightness and neutral infrared color. However, RV measurements over a time baseline of 13 years combined with AO imaging places a lower-limit on the companion mass of $M \geq 0.369 \pm 0.010 M_{\odot}$. We describe the observations acquired from Keck and the Large Binocular Telescope that detect (\S\ref{sec:observations}), confirm (\S\ref{sec:astrometry}), and provide a provisional characterization of the physical properties of the WD (\S\ref{sec:properties}). HD~169889~B adds to a small but growing list of compact objects in the local sample that can be studied in detail by inferring properties from its optical host star and understanding their interaction. 

\begin{figure*}[!t]\label{fig:Doppler}
\begin{center}
\includegraphics[height=3.2in]{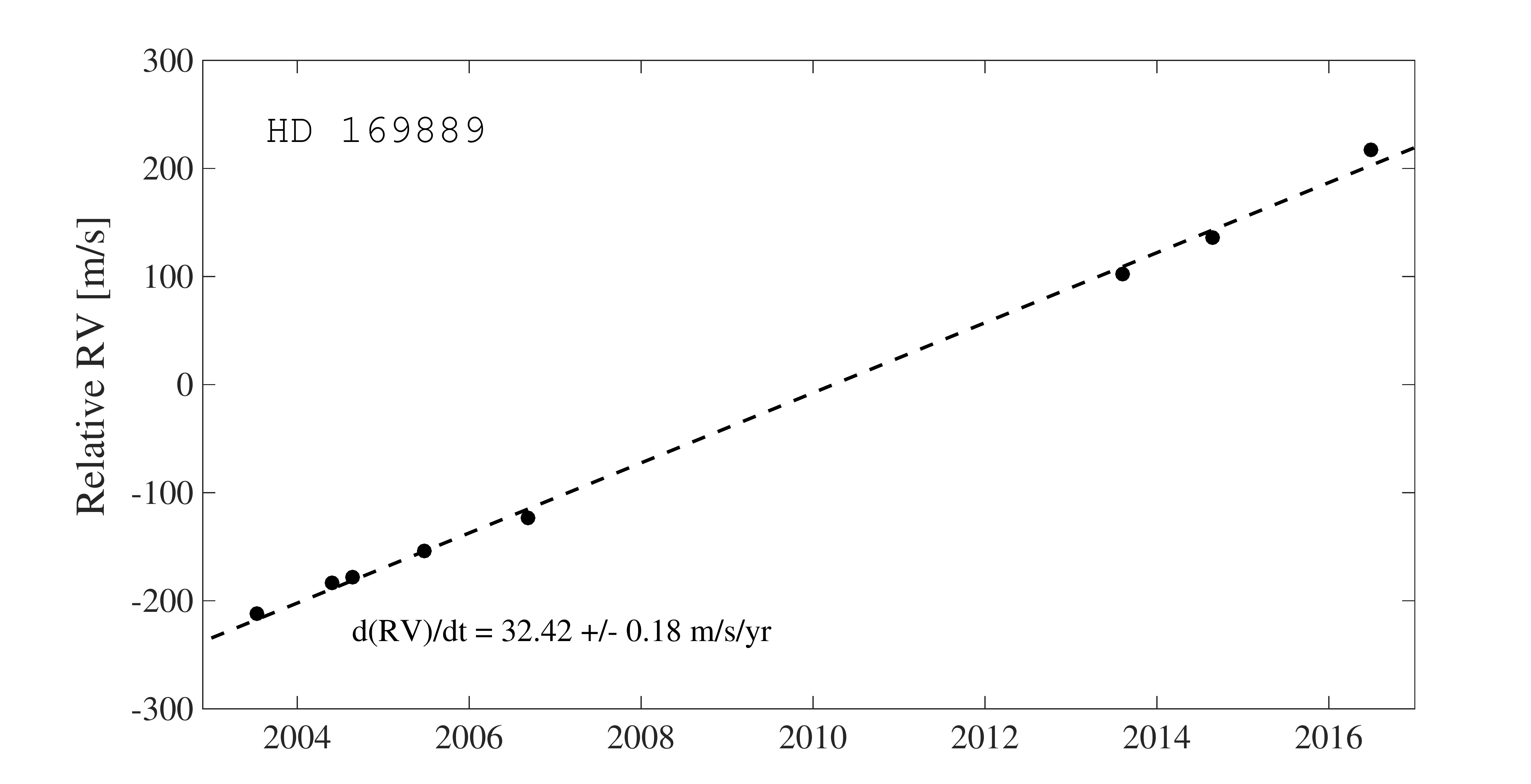} 
\caption{Stellar RV measurements for HD~169889 using HIRES at Keck. A long-term Doppler acceleration indicates the presence of a distant companion. Uncertainties are shown but too small to see at this scale.}
\end{center}
\end{figure*} 

\begin{threeparttable}[!ht]
\centerline{
\begin{tabular}{lc}
\hline
\hline
\multicolumn{2}{c}{HD~169889 Properties}     \\
\hline
\hline
right ascension (J2000)            &   18 26 21.94         \\
declination (J2000)                   &    +08 36 56.74       \\
$B$ (mag)                                 &  9.02         \\
$V$ (mag)                                 &  8.27         \\
$R$ (mag)                                 &  7.8           \\
$I$ (mag)                                  &  7.4            \\
$J$ (mag)                                 &  6.906$^1$ \\
$H$ (mag)                                &  6.560$^1$ \\
$K_s$ (mag)                            &   6.486$^1$  \\
$\pi$ (mas)                               &  $28.297 \pm 0.066$$^2$ \\
d (pc)                                        &  $35.34 \pm 0.08$$^2$   \\
$\mu_{\alpha}$ (mas yr$^{-1}$)            & $-195.671 \pm 0.073$$^2$ \\
$\mu_{\delta}$ (mas yr$^{-1}$)            & $-469.247 \pm 0.066$$^2$ \\
\hline
Mass ($M_{\odot}$)           & $0.89\pm0.13$ \\
Radius ($R_{\odot}$)         &   $0.88\pm0.04$  \\
$\log L$ ($L_{\odot}$)        &  $-0.24\pm0.04$ \\
$\log R'_{HK}$                   &  -4.9        \\
$S_{\rm ave}$                    &  0.19       \\
Gyro Age (Gyr)                   &  $5.2^{+1.3}_{-1.5}$ $^3$ \\
SME Age (Gyr)                   &   5.5-12.9         \\
$\mbox{[Fe/H]}$ (dex)        &   $-0.14\pm0.01$ $^4$    \\
$\log g$ (cm $\mbox{s}^{-2}$) &   $4.49\pm0.03$   $^4$   \\ 
$T_{\rm eff}$ ($K$)             &   $5360 \pm 25$  $^4$   \\
Spectral Type                     &   G9 V                      \\
$v \; \sin(i)$ (km s$^{-1}$)  &   $0.50$                   \\
\hline
	\end{tabular}}
	\begin{tablenotes}
	 \small
	 \vspace{4pt}
	 \item [1] NIR magnitudes from 2-Micron All Sky Survey (2MASS) catalog of point sources \citep{cutri_03,skrutskie_06}. 
	 \item [2] Parallax from {\it Gaia} mission DR2 \citep{gaia_2016,gaia_2018}.
	 \item [3] Gyrochronological age based upon empirical relations \citep{mamajek_hillenbrand_08}.
	 \item [4] Spectral fitting results from \citep{brewer_16}.
	\end{tablenotes}
\caption{Upper panel: Observational parameters of HD~169889~A. Lower panel: Physical properties are derived from HIRES template spectra and theoretical isochrones \citep{valenti_fischer_05}.}
\label{tab:star_props}
\end{threeparttable}

\section{OBSERVATIONS}\label{sec:observations}

\subsection{High-Resolution Spectroscopy}
Precise stellar radial velocity measurements were obtained as part of the California Planet Search program \citep{howard_10a}. HD~169889 was observed using the High Resolution Echelle Spectrometer (HIRES) with the Keck I telescope \citep{vogt_94}. Standard methods for precise Doppler observations were used for spectral calibration, extraction, and RV analysis \citep{howard_10b}. A total of eight Doppler measurements (relative to an arbitrary zero-point) were recorded from 2003 through 2016 (Table~2); a significant acceleration was noticed within the first several years of RV follow-up. HD~169889 was subsequently observed with a relatively low cadence compared to Doppler searches meant specifically for planets. As with other TRENDS discoveries, visual inspection of RV time series measurements along with numerical investigation of Doppler accelerations led to the identification of HD~169889 as a high priority target for follow-up high-contrast imaging \citep{crepp_12a,crepp_14}. 

Fitting the RV slope as a straight line, we find an average acceleration of $\dot{v} = 32.42 \pm 0.18$ m s$^{-1}$ yr$^{-1}$. \footnote{Doppler analysis takes into account the velocity offset in 2004 resulting from the HIRES detector upgrade, which is treated as a nuisance parameter through the statistical (Markov Chain Monte Carlo) marginalization process.} As can be seen in Figure~\ref{fig:Doppler}, when comparing a systemic trend to the HIRES measurements a low-level change in the acceleration over time can be seen. While the instantaneous acceleration places a lower limit on the directly imaged companion mass (see $\S$\ref{sec:mass}), curvature information (the ``jerk", $\ddot{v}$) can ultimately be used to constrain eccentricity and other orbital parameters given sufficient imaging and RV data \citep{crepp_12b}.  

\begin{figure}[!t]\label{fig:discovery}
\begin{center}
\includegraphics[width=5.4in,trim=14.5cm 2cm 0cm 0cm,clip]{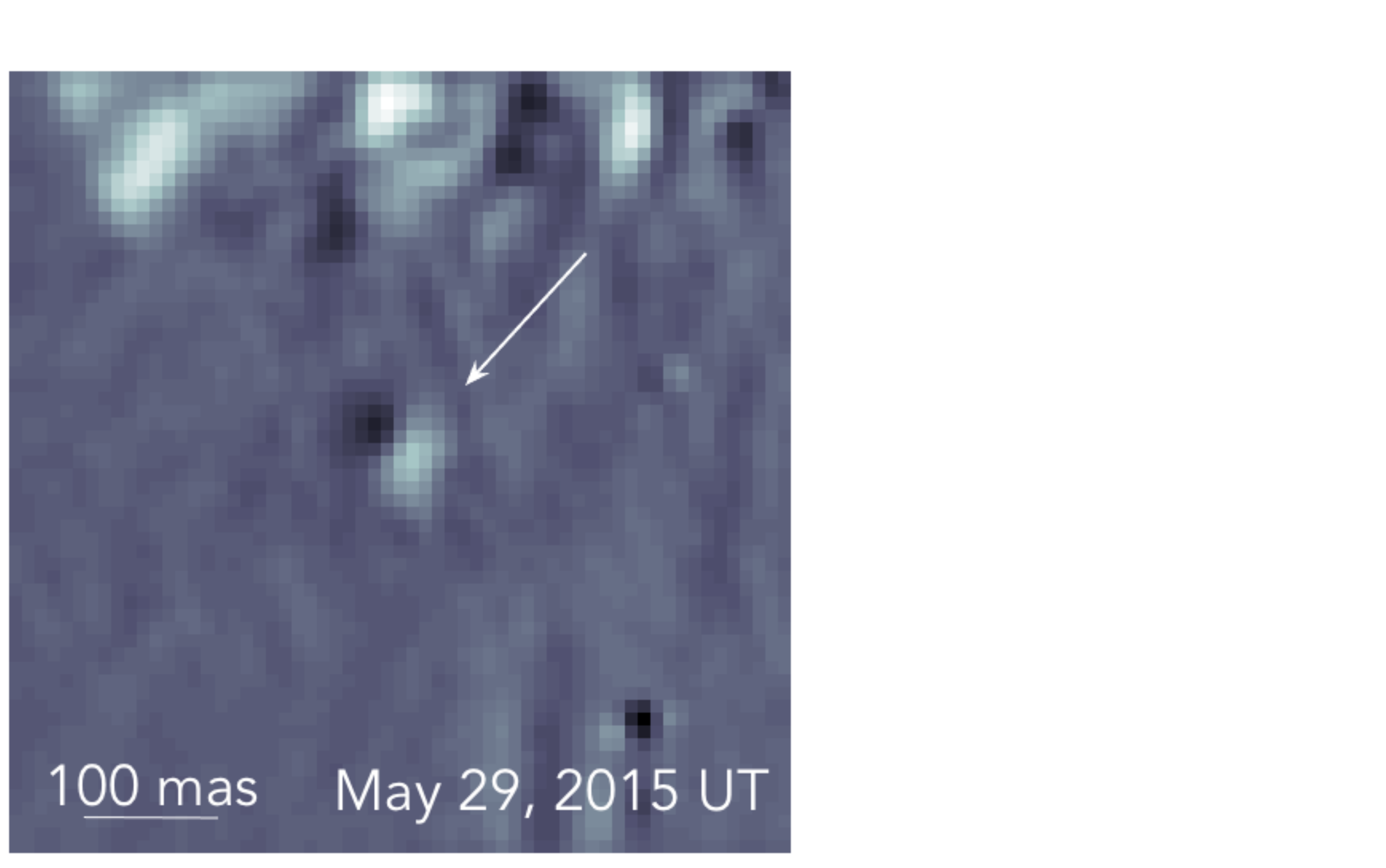} 
\caption{Discovery image of HD~169889~B. Recorded in the $K_s$ filter with field minimal rotation, this image provided sufficient motivation to re-observe the star and ultimately confirm the WD companion.}
\end{center}
\end{figure} 

\begin{figure*}[!t]\label{fig:followup}
\begin{center}
\includegraphics[height=2.2in]{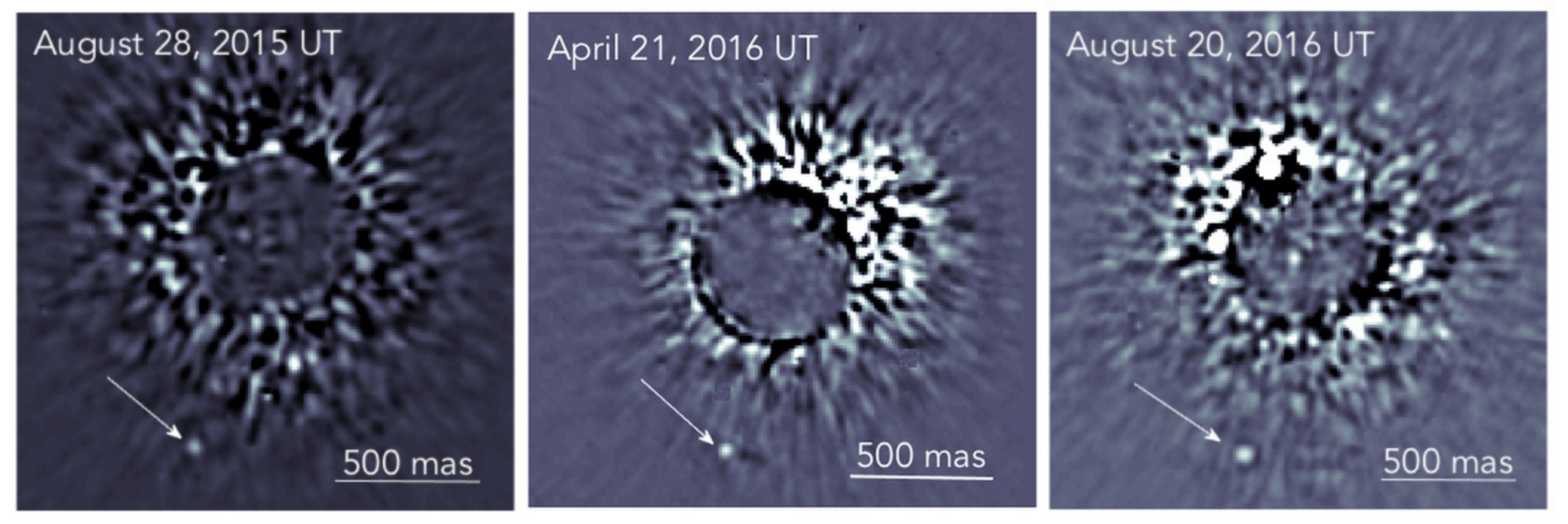} 
\caption{Images of HD~169889~B taken with NIRC2 in the $K_s$-band (left, right) and $H$-band (middle).}
\end{center}
\end{figure*} 

\begin{figure}[!t]\label{fig:lbt}
\begin{center}
\includegraphics[height=3.2in,trim=9.5cm 2.5cm 0cm 0.5cm,clip]{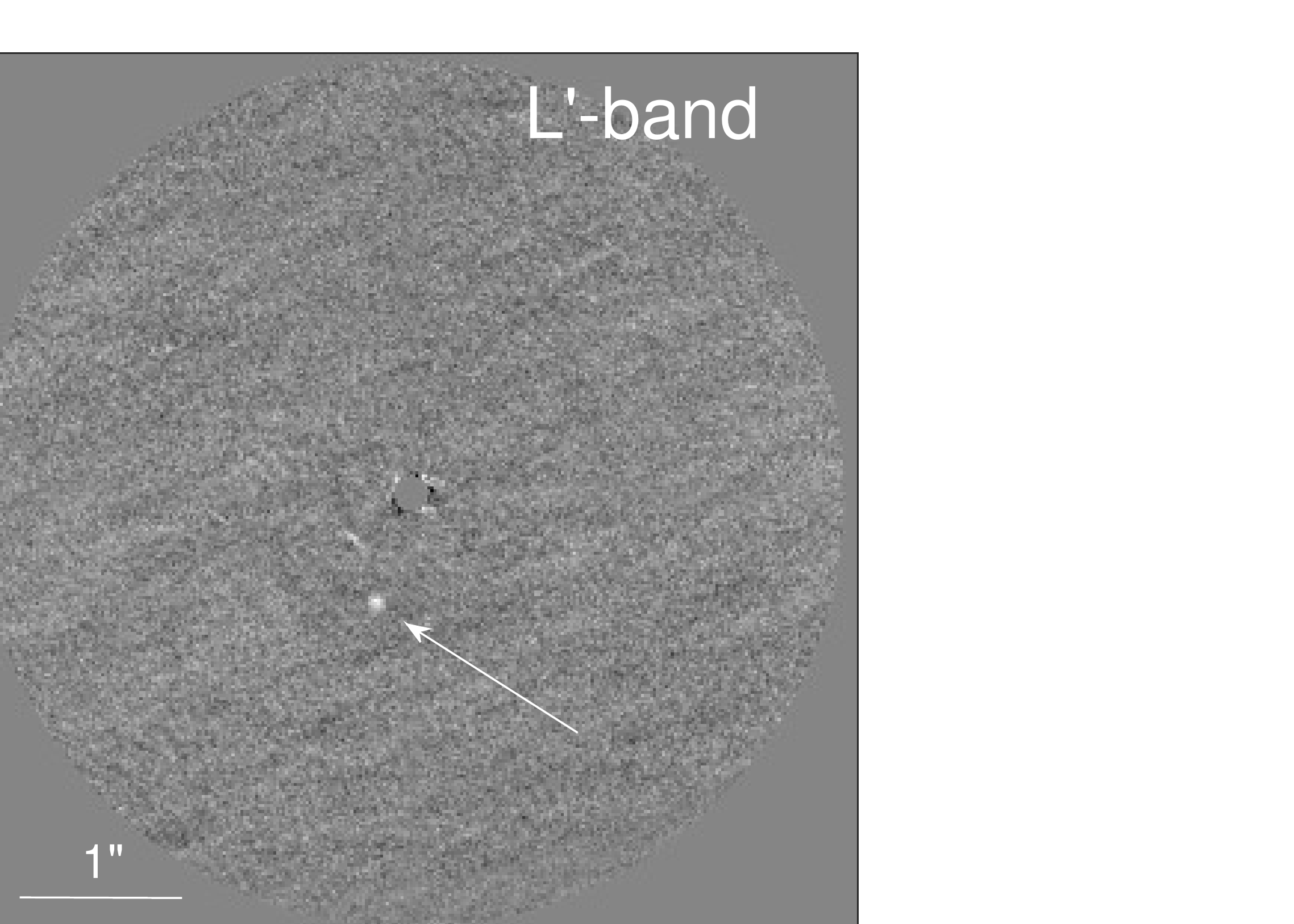} 
\caption{Confirmation image of HD~169889~B taken with LMIRCam at the LBT in June 2016.}
\end{center}
\end{figure} 

\begin{table}[!ht]\label{tab:rvs}
\label{tab:rvs}
\centerline{
\begin{tabular}{cccc}
\hline
\hline
Date                      & BJD                         &   RV                    & Uncertainty  \\
 $\mbox{(UT)}$       &      -2,450,000        &  (m~s$^{-1}$)      &  (m~s$^{-1}$)  \\  
\hline
\hline
2003-07-12  & 2832.8151  & -212.01  &   1.16   \\
2004-05-28  & 3154.0357   & -183.81  &   3.12   \\
    \hline
2004-08-23    & 3240.8538  & -178.44   &  0.84   \\
2005-06-25    & 3546.8841  & -153.44  &   0.81   \\
2006-09-05    & 3983.7572  & -123.22   &  0.80   \\
2013-08-08    & 6512.7585  & 102.17    & 1.16   \\
2014-08-24    & 6893.7523  & 135.61   &  0.90   \\
2016-06-29    & 7568.8750  & 216.75    & 0.88   \\
  \hline
  \hline
\end{tabular}}
\caption{Relative stellar RV measurements for HD~169889. A horizontal line denotes the division between data sets when the HIRES detector was upgraded (August 2004) requiring a relative RV offset for analysis.} 
\end{table}

\subsection{High Contrast Imaging}

\subsubsection{Keck/NIRC2}
HD~169889 was observed using NIRC2 with the Keck II AO system on 2015-May-29 UT, 2015-Aug-28 UT, and 2016-Apr-21 UT \citep{wizinowich_00}. Measurements were obtained at near-infrared wavelengths using the narrow field camera mode ($1024 \times 1024$ pixels with 10 mas plate scale). The field derotator was turned off (vertical angle mode) to enable speckle suppression through angular differential imaging \citep{marois_06}. Table~3 lists a summary of the observations. 

Initially meant to vet the target for binarity with a snap-shot imaging sequence, the first observations of HD~169889 acquired in May 2015 resulted in a marginal detection of what would later be confirmed as the optical companion which we refer to as HD~169889~B. Figure~2 shows the dipole pattern that results from a partially subtracted companion point spread function (PSF) close to the optical host star ($\theta \approx 832$ mas). Given that the first epoch yielded only coarse astrometry and photometry information, subsequent epochs with larger angular rotation were obtained to avoid self-subtraction and distinguish the object from speckle noise. Follow-up observations were obtained in $H$ and $K_s$ with NIRC2 in subsequent observing seasons to assess companion color and relative proper-motion. Due to constraints from weather, parallel observing programs, and other logistics, unsaturated images of the optical host star were only obtained in the $H$-band (April 2016) to allow for relative photometry measurements, making the $K_s$ data (August 2015, August 2016) only helpful for astrometry ($\S$\ref{sec:astrometry}). Standard data analysis methods were used to fully process the differential imaging sequence as employed in other TRENDS detections \citep{crepp_14,crepp_16}. %Mid-infrared observations of the companion were attempted with NIRC2 in April 2016 but resulted in a non-detection due to moderately high background levels and weather. 

\subsubsection{LBT/LMIRCam}
A separate sequence of mid-infrared observations were recorded by LMIRCam at the Large Binocular Telescope (LBT) on June 20 2016 UT \citep{skrutskie_10}. With fewer optical reflections than traditional AO systems, LBT offers lower thermal background levels in the mid-infrared. Imaging measurements were obtained using the LBT Interferometer (LBTI) but using only the left side 8.4m aperture \citep{hinz_16}. Since the LBTI instrument has no de-rotator, all fields are observed to rotate with the parallactic angle on the detector. To track time-variable sky background and detector drifts we nodded the star up and down with a throw of 4.5'' every 50 frames. 

Data were reduced using the LEECH-survey pipeline \citep{skemer_14}. In short, the pipeline implements the following steps. Bad pixels are fixed by replacing their values with the median of the nearest eight functioning pixels. The median of each detector channel is subtracted from the corresponding pixel columns to correct for bias drifts on timescales shorter than nods. Background emission is removed from each image by subtracting the median of 50 images recorded in the opposite nod position taken closest in time. Each image is dewarped using the coefficients reported by \citep{maire_15}. Since LBTI/LMIRCam pixels over sample the single-aperture PSF, we bin each image into $2 \times 2$ pixels, which has the effect of removing any residual bad pixels or cosmic ray hits. Images are registered using a cross-correlation method and then median combined into sets of 20 or sets with less than two degrees of rotation. This rotation limit is chosen so that a companion at the edge of the reduced $3'' \times 3''$ field of view will move by less than $\approx1$ PSF width.

The LEECH pipeline implements principal component analysis (PCA) to fit and remove the influence of the central star before de-rotating and stacking images \citep{soummer_12,amara_quanz_12}. PCA proceeds annulus-by-annulus using a width of 9 pixels ($\approx 2 \lambda/D$) to fit for the starlight and then subtracting the best fit from an annulus only 1-pixel wide. We fit for the optimal number of principal components at each radius by injecting fake planets and iterating until we reach the best contrast. The position and flux of HD 169889 B were fit simultaneously by subtracting a shifted and scaled image of the unsaturated primary star, which we used as a model PSF. This was done for the median combined images before the high-contrast data processing steps to properly account for algorithm effective throughput.

%Email from Jordan Stone (Feb. 5th):
We use the spectral-type--color relations from \citealt{bessell_brett_88} to estimate the $L'$ magnitude of HD~169889~A, finding $L'=6.44$. This result is consistent with WISE photometry, $W_1=6.45\pm0.07$, which is comparable to $L'$ at this level of precision. The resulting apparent magnitude and absolute magnitude of HD~169889~B are listed in Table~\ref{tab:photometry}. 

\begin{table*}[t]\label{tab:obs}
\centerline{
\begin{tabular}{lcccc}
\hline
\hline
Date (UT)       &   Instrument    &    Filter       & $\Delta t_{\rm exp} \times n_{\rm coadds} \times N_{\rm fr}$ & $\Delta \pi$ ($^{\circ}$)  \\
\hline
\hline        
29 May 2015             &   NIRC2           &   $K_s$       &      $5 \times 10 \times 18$     &  1.8    \\
28 August 2015         &   NIRC2           &   $K_s$      &       $5 \times 10 \times 30$     & 48.7    \\
21 April 2016             &    NIRC2          &   $H$         &       $3 \times 15 \times 30$     &  10.5    \\
20 June 2016            &    LMIRCam     &   $L'$         &      $0.08733 \times 1 \times 3948$  &   50   \\
20 August 2016        &     NIRC2          &   $K_s$      &      $10 \times 3 \times 45$     &  20.5    \\
\hline
\hline
\end{tabular}}
\caption{Summary of high-contrast imaging observations showing the number of coadds ($n_{\rm coadds}$), frames ($N_{\rm fr}$), exposure time per frame ($\Delta t_{\rm exp}$), and change in parallactic angle $\Delta \pi$ ($^{\circ}$).}
\label{tab:imaging}
\end{table*}

\begin{table*}[t]\label{tab:astrometry}
\centerline{
\begin{tabular}{lcccc}
\hline
\hline
Date (UT)            &   JD-2,450,000   &      $\rho$ (mas)              &    P. A. ($^{\circ}$)    &     Proj. Sep. (au)    \\
\hline
\hline
29 May 2015       &  7171.9               &    $\approx832\pm10$   &  $\approx163\pm1$  &  $29.4\pm0.4$   \\
28 August 2015   & 7262.8                &   $816\pm6$                  &    $161.7\pm0.8$     &   $28.8\pm0.2$  \\
21 April 2016       & 7500.1                &   $789\pm12$                &    $162.1\pm0.7$     &   $27.9\pm0.4$  \\
20 June 2016       &  7559.9              &    $828\pm3$                 &   $162.0\pm0.5$       &  $29.3\pm0.1$  \\
20 August 2016    &  7620.8              &    $796\pm6$                &   $161.5\pm0.7$       &   $28.1\pm0.2$  \\ 
\hline
\hline
\end{tabular}}
\caption{Summary of astrometry measurements. Position angle (P.A.) measurements are referenced with respect to the equator of the observing epoch.}
\label{tab:imaging}
\end{table*}

\section{Astrometry}\label{sec:astrometry}
\begin{figure*}[!t]\label{fig:confirmation}
\begin{center}
\includegraphics[height=4in]{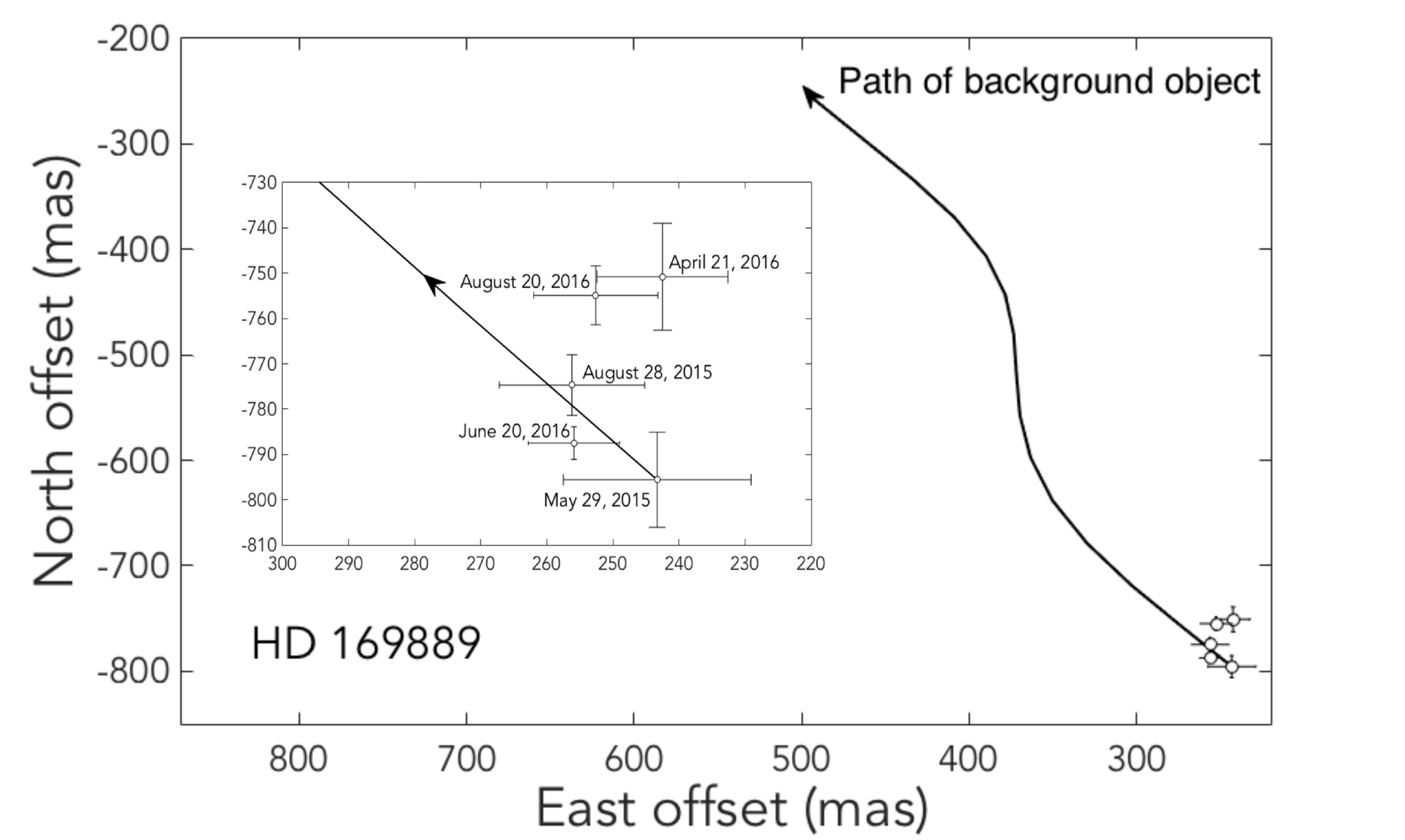} 
\caption{Proper motion and parallactic motion analysis taking into account the path that an infinitely distant background object would traverse is compared to the separation of HD~169889~B over time. Angular offsets are measured relative to the optical primary star. The inset shows a zoomed in view of individual astrometric measurements. The companion shares the same space motion demonstrating that they are gravitationally associated.}
\end{center}
\end{figure*}

HD~169889 has a large proper motion across the sky (Table 1). An astrometric time baseline of 1.2 years (May 2015 to August 2016), which includes five imaging detections, provided unambiguous confirmation that the companion was comoving. We plot the measured position of HD~169889~B relative to HD~169889~A and compare to the null-hypothesis that describes the path of a distant background source (Fig.~\ref{fig:confirmation}).\footnote{NIRC2 was opened for installation of new filters just prior to the first observation of HD~169889~B. The plate scale, image distortion solutions, and position angle offset from \citealt{service_16} are applicable to all measurements.} 

As mentioned previously, the discovery image of HD~169889~B from May 2015 achieved only marginal parallactic angle rotation (Table~3). Nevertheless, we still use astrometry from this epoch. Though the dipole intensity pattern of the companion could not be removed, it can still be modeled by injecting fake companions. Following a procedure similar to that which allows for photometric calibration of self-subtraction, we estimate the position angle (PA) of the companion from this epoch and adopt conservative uncertainties to accommodate systematics from rotational shear (Table~4). All other epochs were analyzed self-consistently using the methods described in other TRENDS discoveries (Crepp et al. 2014; Crepp et al. 2016). 

% We conservatively adopt a $\pm1$ pixel uncertainty for the companion angular separation. 
%bright patch: 161.05$^{\circ}$ 
%middle of dipole: 162.96$^{\circ}$. 

\section{Companion Mass Limit from Dynamics}\label{sec:mass}
We use the precise stellar RV measurements obtained from HIRES along with imaging follow-up measurements to place a lower limit on the mass of HD~169889~B from dynamics. The eight Doppler measurements are fit with a straight line to estimate the acceleration of the star induced by the companion. We augment the internal RV uncertainties listed in Table~2 with a ``jitter" term that is added in quadrature to accommodate the expected level of noise introduced from stellar variability. HD~169889~A has a visual color of $B-V=0.75$ and median activity index $S_{HK}=0.18$. We assume a $B-V$ uncertainty of $0.01$ mags. Using empirical relations from \citealt{isaacson_fischer_10}, we estimate a jitter value of $\sigma=2.2\pm0.1$ m/s. 

A straight line is fit using Markov Chain Monte Carlo (MCMC) methods to find a best-fitting RV slope of $\dot v = 32.42 \pm 0.18$ m s$^{-1}$ yr$^{-1}$. Combined with the direct imaging projected separation, we use the relation from \citealt{liu_02} to derive a lower limit to the companion mass of $m_{\rm min}=0.369 \pm 0.010 M_{\odot}$. Clearly, HD~169889~B cannot be a brown dwarf but is instead a compact object provided the acceleration is attributable to this object. 

A more careful treatment would consider the slight curvature noticeable by eye in the RV time series as well as astrometry (Figure~1). However, only a small fraction of the companion orbit has been traced in either data set. Further, the astrometry measurements were obtained with two different instruments using three different filters. The systematics that result from combining these observations can be noticed in the zoomed in portion of Figure~5 at the $\approx 1\sigma$ level (10 mas). The WD companion appears to be moving to the north, i.e. closer to HD~169889~A, but further AO follow-up imaging is required to confirm. 

% Cannot get true physical separation because we don't yet know mass of WD. 
% Since we have a lower limit on the now the mass of the WD, we calculate a lower limit to the true physical separation (Howard et al. 2010). Can we really do this?  Could we use the primary star mass estimate to perform this calculation? No. Would need to know the radial acceleration of the WD.  Rather than a comprehensive dynamical analysis, we instead estimate the companion true physical separation based on the angular separation, parallax, and radial acceleration. Using the method outlined in Howard et al. 2010 and implemented for other TRENDS studies (Crepp et al. 2014; Crepp et al. 2016), we ... This in turn places a lower limit on the orbital period corroborating the notion that the WD has a long orbital period. As done with other TRENDS detections, we use the method from Howard et al. to estimate a physical separation of ... au. Period lower limit of ... years. \\

\begin{table}[!ht]\label{tab:photometry}
\centerline{
\begin{tabular}{lc}
\hline
\hline
\multicolumn{2}{c}{HD~169889~B}     \\
\hline
\hline
$\Delta H$                   &   $9.76\pm0.16$  \\
$\Delta L'$                   &   $9.60\pm0.03$   \\
$H$                              &   $16.32\pm0.16$   \\
$L'$                               &   $16.05\pm0.08$   \\
H-L'                               &    $0.27 \pm 0.18$  \\
$M_H$                          &   $13.59\pm0.16$   \\
$M_{L'}$                       &     $13.32\pm0.08$    \\
$M_{\rm dyn}$ ($M_{\odot}$)   &  $\geq 0.369 \pm 0.010$   \\
$t_{\rm cooling}$ (Gyr)            &   $\leq  5.2^{+1.3}_{-1.5}$    \\
\hline
\end{tabular}}
\caption{Photometric results and companion properties.} 
\label{tab:comp_props}
\end{table}

%Photometry: we measure H-band relative flux using two methods: (1) injected fake companions (Explain) and unocculted frames of stellar psf; (2) direct transmission through semi-transparent coronagraphic mask. Could not get these to agree with one another. I think star may be too close to edge of mask. 

%See notebook for an estimate of dK. We find approximately $\Delta K \approx 9.71$, which is consistent with the above but does not account for self-subtraction. 

\section{White Dwarf Properties}\label{sec:properties}

Empirical relations between WD masses and their progenitors have been established based on the study of open clusters, globular clusters, and common proper motion pairs \citep{weidemann_00,williams_09}. Using the dynamical mass lower limit from RVs and imaging, we attempt to place a lower limit on the companion progenitor mass ($M_i$) by inverting the empirical relationship from \citealt{catalan_08},
\begin{equation}
M_f = (0.096 \pm 0.005)M_i + (0.429 \pm 0.015),
\end{equation}
which is applicable for $M_i \leq 2.7M_{\odot}$ stars.\footnote{Both $M_i$ and $M_f$ are implicitly listed in solar mass units.} We find however that the dynamical mass lower limit is smaller than all final mass ($M_f$) values used to establish the empirical relation. Naively extrapolating the equation to smaller masses is not (justifiable nor) informative since the corresponding progenitor mass ($M_i$) becomes smaller than the mass of HD~169889~A ($M_i \leq 0.89 M_{\odot}$), which cannot be true provided the pair formed at the same time. 
%We ignore metallicity effects as they are expected to be much smaller than the current uncertainty in WD dynamical mass. 

We instead place a constraint on the nuclear-burning plus cooling age of the WD by estimating the age of HD~169889~A.\footnote{Progenitor mass estimates are further complicated by chemical composition, e.g. whether the atmosphere is hydrogen dominated, helium rich, or a combination thereof.} Using gyrochronology relations from \citet{mamajek_hillenbrand_08}, we find $P_{\rm rot}=33\pm4$ days, corresponding to a gyrochronological age of $\tau = 5.2^{+1.3}_{-1.5}$ Gyr \citep{wright_04}. This estimate, which is based on the $B-V=0.75$ color of the primary (we assume $\pm0.01$ uncertainty) and $R'_{HK}=-4.9$ value (we assume $\pm0.1$ uncertainty), is consistent with the broad range suggested from an isochronal analysis (SME, Table~1). 

To break the degeneracy between nuclear-burning timescale and cooling age it is necessary to precisely determine the WD effective temperature \citep{liebert_05}. With only two broadband photometry measurements available, we consider the Rayleigh Jeans tail of the Planck curve to get a handle on the WD effective temperature. Using the $H$ and $L'$ measurements to obtain SED slope information, $dB_{\rm \lambda}/d \lambda \approx (B_{L'}-B_{H})/ \Delta \lambda$, we find a temperature of $T \approx 2150$ K. This value however would imply a black body peak in the near-infrared ($\lambda_0 \approx 1.4 \; \mu$m) and is therefore inconsistent with the presumption that the WD is hot in the first place. We adopt $T=2150$ K as a lower-limit to the companion temperature, noting that WD's this cool have not been detected. 

Knowing a priori that ultra-cool WDs are scarce, we fit a full black body curve to the SED by incorporating the distance to the source. We find that hotter temperatures, $T \approx 10,000$ K, are also consistent with the limited photometry available; however, large discrepancies still remain in the mid-infrared suggesting that the presence of Hydrogen and Helium may need to be accounted for to properly fit the data set. 

Finally, we use the H-band absolute magnitude of HD~169889~B and compare to WD cooling models (see \citealt{holberg_06,kowalski_06,tremblay_11,bergeron_11}).\footnote{\url{http://www.astro.umontreal.ca/~bergeron/CoolingModels}} Masses in the range from $M=0.4-1.2M_{\odot}$ are considered. The mass lower-limit is matched to the lower-limit from dynamics, and the mass upper limit corresponds to that available from theoretical evolutionary sequences. In this range, pure Hydrogen atmospheres result in an effective temperature range of $\approx$4,000-12,000 K with a cooling age of 5.3-1.9 Gyr, and pure Helium atmospheres result in an effective temperature range of $\approx$4,000-14,000 K with a cooling age of 5.8-1.3 Gyr. Given the gyrochronological age estimate of HD~169889~A of $\tau=5.2^{+1.3}_{-1.5}$ Gyr, we cannot at this point identify any tension with the models without further refinements to the companion mass. While these estimates are consistent with the black-body temperature approximated using the H-L color, low resolution spectroscopy is needed to further characterize the WD and understand its composition, cooling age, and progenitor mass.

\section{Concluding Remarks}\label{sec:conclusions}

% What have we presented? % Why is this important? And what is unique about HD 169889 B? 
We have detected a WD companion that orbits a bright ($V=8.27$) and well characterized solar-type (G9 V) star with a precisely determined parallax ($28.297 \pm 0.066$ mas). The census of these ``Sirius-like" systems is known to be incomplete just beyond the solar neighborhood ($d \approx 25$). The second WD discovery of the TRENDS high-contrast imaging program, HD~169889~B offers an up-close view of a directly imaged compact object for which dynamical mass information is available. 

% What further work should be performed? 
We have estimated the nuclear-burning timescale plus cooling age of the WD using gyrochronology of HD~169889~A. At present however degeneracies between effective temperature, composition, and the companion final mass preclude an estimate of the compact object progenitor mass. Broader SED information along with continued Doppler and astrometric monitoring is warranted to further characterize the companion atmosphere and other physical properties. Space-based observations could provide complementary photometric and spectroscopic measurements and potentially allow for an estimate of log(g) through the Balmer line profile, and a gravitational red-shift measurement would yield the WD mass-to-radius ratio provided contamination from the bright optical host star can be mitigated. It may be worth investigating indirect evidence for past mass transfer from HD~169889~``B" to its optical host star through studies of enhanced metallicity in the optical spectrum. Finally, the companion will be valuable for characterizing high-contrast imaging spectrographs given its flux ratio and presumed featureless spectrum.

% Additional notes: 
% Collissionally induced (hydrogen?) absorption giving rise to non-blackbody colors in cool WD atmospheres (ref). Fraction of hydrogen and helium mixing matters.  
% Cool WD's are relatively rare (ref. Leggett et al. 2011 ... read the first few pages). corresponding to long cooling time. 
% Any chance of mass exchange? How close would they have to be? 
% From Liebert et al. 2005: "An accurate Teff determination is essential for measuring the cooling age of Sirius B." 

\section{ACKNOWLEDGEMENTS}
The TRENDS high-contrast imaging program is supported in part by NASA Origins of Solar Systems grant NNX13AB03G. JRC acknowledges support from the NASA Early Career Fellowship and NSF Career Fellowship. EJG acknowledges support from the NSF graduate research fellowship program. This research has made use of the SIMBAD database, operated at CDS, Strasbourg, France. Data presented herein were obtained at the W.M. Keck Observatory, which is operated as a scientific partnership among the California Institute of Technology, the University of California and the National Aeronautics and Space Administration. The Observatory was made possible by the generous financial support of the W.M. Keck Foundation. The LBT is an international collaboration among institutions in the United States, Italy and Germany. LBT Corporation partners are: The University of Arizona on behalf of the Arizona university system; Istituto Nazionale di Astrofisica, Italy; LBT Beteiligungsgesellschaft, Germany, representing the Max-Planck Society, the Astrophysical Institute Potsdam, and Heidelberg University; The Ohio State University, and The Research Corporation, on behalf of The University of Notre Dame, University of Minnesota and University of Virginia. This work has made use of data from the European Space Agency (ESA) mission {\it Gaia} (\url{https://www.cosmos.esa.int/gaia}), processed by the {\it Gaia} Data Processing and Analysis Consortium (DPAC, \url{https://www.cosmos.esa.int/web/gaia/dpac/consortium}). Funding for the DPAC has been provided by national institutions, in particular the institutions participating in the {\it Gaia} Multilateral Agreement. This paper made use of WD cooling models from \url{http://www.astro.umontreal.ca/~bergeron/CoolingModels}. We are deeply grateful for the vision and support of the Potenziani and Wolfe families.

\end{document}